\documentclass[aps,showpacs,nofootinbib,twocolumn,preprintnumbers,nobalancelastpage]{revtex4}

\usepackage[english]{babel}
\usepackage{amsmath,amssymb}
\usepackage[dvips]{graphicx}
\usepackage[sort&compress]{natbib}
\usepackage{amsfonts}
\usepackage{bbm}
\usepackage{color}

\DeclareMathAlphabet{\mathsc}{OT1}{cmr}{m}{sc}

\allowdisplaybreaks

\newcommand{\ie} {{\it i.e.}}

\def\321{SU(3) $\otimes$ SU(2) $\otimes$ U(1)}

\def\lsim{\raise0.3ex\hbox{$\;<$\kern-0.75em\raise-1.1ex\hbox{$\sim\;$}}}
\def\gsim{\raise0.3ex\hbox{$\;>$\kern-0.75em\raise-1.1ex\hbox{$\sim\;$}}}

\newcommand{\flux}[2][]{\ensuremath{\ifthenelse{\equal{#1}{}}{}{^{#1}\!}\mathit{#2}}}

\newcommand{\sla}[1]{/\!\!\!\!#1}

\begin{document}
\preprint{YITP-SB-11-42}

\title{Present Bounds on New Neutral Vector Resonances from Electroweak 
Gauge Boson Pair Production at the LHC}
\author{O.\ J.\ P.\ \'Eboli}
\email{eboli@fma.if.usp.br}
\affiliation{Instituto de F\'{\i}sica,
             Universidade de S\~ao Paulo, S\~ao Paulo -- SP, Brazil.}

\author{J.\ Gonzalez--Fraile}
\email{fraile@ecm.ub.es}
\affiliation{%
  Departament d'Estructura i Constituents de la Mat\`eria and
  ICC-UB, Universitat de Barcelona, 647 Diagonal, E-08028 Barcelona,
  Spain}
\author{M.\ C.\ Gonzalez--Garcia} \email{concha@insti.physics.sunysb.edu}
\affiliation{%
  Instituci\'o Catalana de Recerca i Estudis Avan\c{c}ats (ICREA),}
\affiliation {Departament d'Estructura i Constituents de la Mat\`eria, Universitat
  de Barcelona, 647 Diagonal, E-08028 Barcelona, Spain}
\affiliation{%
  C.N.~Yang Institute for Theoretical Physics, SUNY at Stony Brook,
  Stony Brook, NY 11794-3840, USA}

\begin{abstract}

  Several extensions of the Standard Model predict the existence of
  new neutral spin--1 resonances associated to the electroweak
  symmetry breaking sector. Using the data from ATLAS (with integrated
  luminosity of $\mathcal{L}=1.02\ \mbox{fb}^{-1}$) and CMS (with
  integrated luminosity of $\mathcal{L}=1.55\ \mbox{fb}^{-1}$) on the
  production of $W^+W^-$ pairs through the process $pp \to \ell^+
  \ell^{\prime -} \, \sla{E}_T$, we place model independent bounds on
  these new vector resonances masses, couplings and widths. Our analyses show
  that the present data excludes new neutral vector resonances with
  masses up to 1--2.3 TeV depending on their couplings and widths. We
  also demonstrate how to extend our analysis framework to
  different models working a specific example.

\end{abstract}


\maketitle


\section{Introduction}

One of the primary physics goals of the CERN Large Hadron Collider
(LHC) is the direct study of the electroweak symmetry breaking (EWSB)
sector via the production of new states associated to it.  The
analyses of unitarity in the weak gauge boson scattering $W^+_L W^-_L
\to W^+_L W^-_L$ indicates that there must be a contribution of the
EWSB at the TeV scale~\cite{Lee:1977eg}, well within the LHC
reach. There is a plethora of possibilities for the EWSB sector that
contains new scalar and vector resonances, and the Standard Model (SM)
represents only the minimal scenario, with a Higgs sector with one
scalar Higgs boson being responsible for cutting off the growth of the
weak gauge boson scattering amplitudes. 
\smallskip

New vector resonances are a common feature of models where the EWSB is
due to a new strongly interacting sector~\cite{TC}.  Although the
precision electroweak measurements and flavor changing neutral
currents present an obstacle for strongly interacting theories, recent
theoretical advances made possible the construction of models in
agreement with the experimental constraints~\cite{NTC}. Furthermore,
new spin--1 states are also present in extra dimension scenarios, in
particular in Higgsless models~\cite{hless} where unitarity
restoration takes place through the exchange of an infinite tower of
spin--1 Kaluza-Klein excitations of the known electroweak gauge
bosons~\cite{Csaki:2003dt}. Such scenarios can be viewed as the
holographic version of strongly coupled theories~\cite{ads-cft}.
\smallskip

In this work, we derive bounds on new neutral spin--1 resonances
($Z^\prime$) associated to the EWSB from the available ATLAS and CMS
data on $W^+W^-$ pair production
\begin{equation}
  p p \to Z^\prime \to W^+ W^- \to \ell^+ \ell^{\prime -} \, \sla{E}_T
\label{eq:proc}
\end{equation}
where $\ell$ and $\ell^\prime$ stand for electrons and muons.  We
perform a model independent analysis proposed in
Refs.~\cite{Alves:2009aa,Eboli:2011bq}.  We present our results as
constraints on the relevant spin-1 boson effective couplings, mass and
width. For instance, our results indicate that $Z^\prime$'s coupling
with SM strength to light quarks and to pairs $W^+W^-$ saturating the
partial wave amplitudes can be excluded at 95\% CL if their masses are
lighter than $\simeq 1750$ GeV. \smallskip

This paper is organized as follows. In Section \ref{sec:frame} we
present our model independent parameterization of the $Z^\prime$
properties.  Section~\ref{ana:frame} contains a detailed accounting of
the procedures used in our analyses.  Our model independent results
are presented in Section~\ref{sec:results} while we show in
Section~\ref{sec:mdr} that our analysis framework can be
adapted to a specific model.  Our conclusions are drawn in
Section~\ref{sec:summary}. \smallskip

\section{Parameterization of the $Z^\prime$ properties}
\label{sec:frame}

In order to evaluate the $Z^\prime$ production cross section via the
channel (\ref{eq:proc}) we must know the $Z^\prime$ couplings to light
quarks and $W^+W^-$ pairs in addition to its mass and width.  We do
not assume any relation between these parameters (although they might
be connected in a complete theory). Nevertheless, inspired by models
where the new vector states interact with the light quarks and
electroweak gauge boson via their mixing with the SM vectors, we
assume that the $Z^\prime$ couplings exhibit the same Lorentz
structure as those of the SM. 
\smallskip

We normalize the $Z^\prime W^+ W^-$ coupling by the value
${g_{Z^\prime WW}}_{max}$ that saturates the partial wave amplitude
for the process $W^+ W^- \to W^+ W^-$ by the exchange of a $Z^\prime$,
~\cite{Birkedal:2004au}, {\em i.e.}
\begin{equation}
{g_{Z^\prime WW}}_{max}=g_{ZWW}\, \frac{M_Z}{\sqrt{3}M_{Z^\prime}} 
\label{eq:gwwvmax}
\end{equation}
where $g_{ZWW}=g~ c_W$ is the strength of the SM triple gauge boson
coupling.  Here $g$ stands for the $SU(2)_L$ coupling constant and
$c_W$ is the cosine of the weak mixing angle. 
\smallskip

We treat the $Z^\prime$ width as a free parameter since it can receive
contributions from particles that do not play a role in our study,
such as $b$ and $t$ quarks. The only bound to the $Z^\prime$ width is
that it should be compatible with its couplings to light quarks and
$WW$ pairs that is expressed by the lower bound ~\cite{Alves:2009aa}
\begin{eqnarray}
&&\Gamma_{Z^\prime}\; >\;0.27\, |G| \, 
\,\left(\frac{M_{Z^\prime}}{M_Z}\right)^2 
{\rm GeV} \; ,
\label{eq:zcouplimit}
\end{eqnarray}
where we have defined the combination 
\begin{equation}
G=\left(\frac{g_{Z^\prime q\bar q}}{g_{Zq\bar q}} \right)\,
\left(\frac{g_{Z^\prime WW}}{{g_{Z^\prime WW}}_{max}}\right)\, ,
\label{eq:G}
\end{equation}
with $g_{Z^\prime q\bar q}$ being the $Z^\prime$ coupling to light quark
pairs and $g_{Zq\bar q}=g/c_W$. 
\smallskip

Within our approach we can express the cross section for the process
(\ref{eq:proc}) as
\begin{equation}
\sigma_{\rm tot}= \sigma_{\rm SM}\, +\, 
G 
\, \sigma_{\rm int}(M_{Z^\prime},\Gamma_{Z^\prime})
\,+\,G^2 \,
\sigma_{Z^\prime}(M_{Z^\prime},\Gamma_{Z^\prime}) 
\label{eq:sigmatot}
\end{equation}
where the Standard Model, interference and new resonance contributions
are labeled SM, int and $Z^\prime$ respectively.  
\smallskip

\section{Analyses framework}
\label{ana:frame}

ATLAS~\cite{ATLASww} and CMS~\cite{CMSww} analyzed the $W^+W^-$ production through the final state
given in Eq.~\eqref{eq:proc}.  Our strategy is to use the SM
backgrounds that have been carefully evaluated by the experimental
collaborations and we simulate only the $Z^\prime$ signal. However, in
order to tune and validate our Monte Carlo we also simulated the SM
production of $W^+W^-$ pairs and compared with the results presented
by ATLAS and CMS. 
\smallskip

We evaluated the signal and SM $W^+ W^-$ cross sections by two
different methods. In the first one, we used the package
MADEVENT~\cite{madevent} to evaluate the ${\cal O}(\alpha^4)$ signal
matrix elements for the subprocesses $q \bar{q} \to \ell^+ \nu
\ell^{\prime -} \nu^\prime$, with $\ell/\ell^\prime = e,\mu$ as well
as the small contribution with $\ell/\ell^\prime=\tau$ which then
decays leptonically into either $e$ or $\mu$ and the corresponding
neutrinos.  Its output is fed into PYTHIA~\cite{Sjostrand:2006za} for
parton shower and hadronization and a simple detector simulation
provided by PGS 4~\cite{pgs4}.  In what follows we will label
it as ``ME+Pythia+PGS-MC''.  A second evaluation was made with a {\sl
  homemade} Monte Carlo that evaluates the process \eqref{eq:proc} at
parton level using the ${\cal O}(\alpha^4)$ signal matrix elements for
the subprocesses $q \bar{q} \to \ell^+ \nu \ell^{\prime -}
\nu^\prime$, with $\ell/\ell^\prime = e,\mu$.  The scattering
amplitudes for the relevant subprocesses were obtained using the
package MADGRAPH~\cite{madevent}.  In what follows we will label this
calculation as ``OUR ME-MC''.  In both cases we used CTEQ6L parton
distribution functions \cite{CTEQ6} and the MADEVENT default
renormalization and factorization scales.

\medskip
\noindent{\bf ATLAS analysis}
\medskip
\begin{figure}
\includegraphics[width=0.49\textwidth]{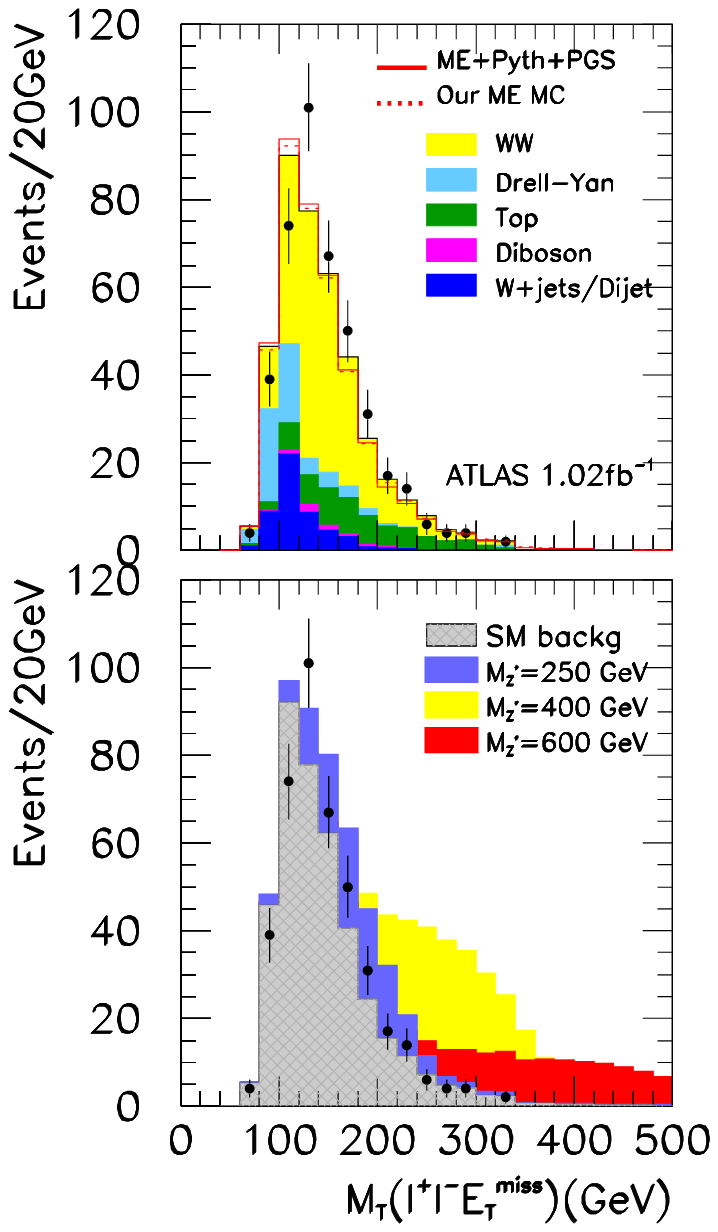}
\caption{Upper panel: Transverse mass distribution of the SM
  contributions to the process $pp \to \ell^+ \ell^{\prime
    -}\sla{E}_T$ calculated by ATLAS (colored histograms) together
  with the number of observed events by ATLAS (points with error bars)
  and the performance of ME+Pythia+PGS-MC (red solid) and OUR ME-MC
  (red dotted).  The results shown correspond to an integrated luminosity
  of $\mathcal{L}=1.02\ \mbox{fb}^{-1}$.
  \\
  Lower panel: Transverse mass distribution of the total SM
  contribution to the process $pp \to \ell^+ \ell^{\prime -}\sla{E}_T$
  (gray hashed) together with the total expected number of events
  including a $Z^\prime$ of 250 GeV with $G=0.5$ (blue), a $Z^\prime$
  of 400 GeV with $G=1$ (yellow) and a $Z^\prime$ of 600 GeV with $G=1$
  (red).  For the three masses $\Gamma_{Z^\prime}=0.06M_{Z^\prime}$.
  We include also the ATLAS observed spectrum.}
\label{fig:dist_atlas}
\end{figure}

The ATLAS simulation of the $W^+W^-$ process was carried out at NLO
and with an accurate detector simulation. In order to take into account
some of these features included in the ATLAS evaluation of the SM $W^+
W^-$ production we normalize our total cross section for the $ee$,
$e\mu$ and $\mu\mu$ channels by an overall factor such that our two
simulations yield the result presented in Table 2 of
Ref.~\cite{ATLASww} after the same cuts have been implemented.  In
particular electrons and muons are accepted if
\begin{eqnarray}
&& |\eta_e|<1.37 \hbox{\ or\ }
1.52<|\eta_e|<2.47\hbox{\ and\ }|\eta_\mu|<2.4 .  
\label{cuts:atlas1}
\end{eqnarray}
Also, the lepton isolation requirement in  ME+Pythia+PGS-MC simulation
is that the sum of the energy in the calorimeter
cells within a cone  $\Delta R <0.3$  around the electron 
must be less than 4 GeV while in a cone $\Delta R <0.2$ around the muon, 
the sum $p_T$ of all other tracks is less than 10\% of the $p_T$ of the muon.
To implement this requirement in OUR ME-MC we simply impose: 
\begin{eqnarray}
&& \Delta R_{ee}>0.3\hbox{\ and\ }\Delta R_{e\mu,\mu\mu}>0.2 \;.
\label{cuts:atlas2}
\end{eqnarray}
Events are selected if they verify that the leading electron in the 
$e^+e^-$ channel  and the electron in the $e\mu$ channel accomplish:
\begin{eqnarray}
&& p_T > 25 \hbox{ GeV,}
\label{cuts:atlas3}
\end{eqnarray}
while for the  muons and the subleading electron in the $e^+e^-$ channel
\begin{eqnarray}
&& p_T > 20 \hbox{ GeV.}
\label{cuts:atlas4}
\end{eqnarray}
Furthermore, 
\begin{eqnarray}
&& M_{\ell\ell} > 15 \hbox{ GeV } \;\;,\;\; M_{e\mu} > 10 \hbox{ GeV},
\nonumber
\\
&& | M_{\ell\ell} -M_Z| > 15 \hbox{ GeV, } 
\label{cuts:atlas5}
\\
&& E_{T,~rel}^{miss}(ee)>40\hbox{ GeV } \;\;,\;\;E_{T,~rel}^{miss}(\mu\mu)>45\hbox{ GeV }
\nonumber\\
&& \hbox{and   }
E_{T,~rel}^{miss}(e\mu)>25\hbox{ GeV }, 
\nonumber
\end{eqnarray}
where $M_{\ell\ell}$ stands for the invariant mass of the lepton pair
and the relative missing energy is defined as:
\begin{equation}
E_{T,~rel}^{miss} =\left\{\begin{array}{clc}
&E_T^{miss}\times\sin{\Delta\phi_{\ell,j}}\ &\hbox{if}\ \ \Delta\phi_{\ell,j}<\pi/2\\
		     &E_T^{miss}\ &\hbox{if}\ \ \Delta\phi_{\ell,j}>\pi/2
                     \end{array} \right.
\label{etmissrel}
\end{equation}
with $\Delta\phi_{\ell,j}$ being the difference in the azimuthal 
angle $\phi$ between the transverse missing energy 
and the nearest lepton or jet.

Finally  in  ME+Pythia+PGS-MC simulation
jets are reconstructed with the anti-$k_T$ algorithm~\cite{arXiv:0802.1189} with a jet resolution 
parameter $\Delta R=0.4$ and we veto events containing jets with
\begin{eqnarray}
&&  p_T>30 \hbox{ GeV  and  } |\eta_j|<4.5 \; .
\label{cuts:atlas6}
\end{eqnarray}

We present in Table~\ref{norma} the overall normalization needed to
tune our simulations to the ATLAS one.  We have also verified
that the relative event reduction due to each cut
\eqref{cuts:atlas3}--\eqref{cuts:atlas5} in our simulations is in
agreement to that reported in Table 2 of Ref.~\cite{ATLASww}.
\smallskip

\begin{table}
\begin{tabular}{|c|c|c|c|c|}
\hline
Experiment & Monte Carlo & $ee$   & $e\mu$   & $\mu\mu$ 
\\
\hline
ATLAS  & OUR ME-MC & 0.54 & 0.78 & 1.04
\\
\hline
ATLAS  & ME+Pythia+PGS-MC & 0.66 & 0.95 & 1.2
\\
\hline
CMS  & OUR ME-MC & 0.50 & 0.73 & 0.84
\\
\hline
CMS  & ME+Pythia+PGS-MC & 0.60 & 0.91 & 1.08
\\
\hline
\end{tabular}
\caption{Overall multiplicative factors used to tune our Monte Carlos
  to the total number of events in the different flavour channels 
  predicted by the ATLAS and CMS simulations.}
\label{norma}
\end{table}


In order to validate our Monte Carlo simulations for the SM $W^+W^-$
production we compare them with the ATLAS prediction for the
transverse mass ($M_T$) spectrum after cuts in the top panel of
Fig.~\ref{fig:dist_atlas}. The results shown corresponds to an
integrated luminosity of $\mathcal{L}=1.02\ \mbox{fb}^{-1}$.  In this
figure we evaluated just the SM $W^+W^-$ production and added the
ATLAS results for the backgrounds.  As we can see, both
ME+Pythia+PGS-MC and OUR ME-MC simulations approximate very well the
ATLAS results.  However, it should be noticed that the three
simulations, the one by ATLAS and two by us, present some discrepancy
with the data at small transverse masses.  
\smallskip

In the simulation of the $Z^\prime$ signal we employed the same
normalization factors obtained from the $W^+W^-$ SM production for the
channels $ee$, $e\mu$, and $\mu\mu$; see Table \ref{norma}. Moreover,
since our two simulations present a similar performance we adopted OUR
ME-MC for our signal calculations because it is much faster. However
we also verified that the results obtained are in agreement with those
from ME+Pythia+PGS-MC for a few points of the parameter
space. 
\smallskip

We present, as an illustration, in the lower panel of Figure
~\ref{fig:dist_atlas} the expected $M_T$ distribution for three
different $Z^\prime$ masses for an integrated luminosity of 1.02
fb$^{-1}$, as reported by ATLAS, and after applying the cuts
\eqref{cuts:atlas1}--\eqref{cuts:atlas6}. The existence of this
neutral vector resonance is characterized by an excess of events at
higher $M_T$ values with respect to the SM expectations. 
\smallskip

Consequently one can use the transverse mass spectrum to place
constraints on the $Z^\prime$ properties.  In order to do so we have
constructed a binned log-likelihood function based on the contents of
the different bins in the transverse mass distribution, {\em i.e.},
the observed number of events $N_d^i$, and the expected events in the
SM, $N_B^i$, plus the expected number of events in the presence of the
$Z'$, $N^i_S$, after applying the cuts
(\ref{cuts:atlas1})--(\ref{cuts:atlas6}). Assuming independent Poisson
distributed $N_d^i$ it reads:
\begin{eqnarray}
&&
-2\ln L_{\rm ATLAS}(M_{Z'},G,\Gamma_{Z'})
=
\nonumber \\
&&\begin{array}{c}\\ [-0.2cm]
{\rm Min}\\[-0.2cm] \xi_j
\end{array} 
\Big\{2\sum_{i=1}^{N^{max}_{AT}} \left [ N^i_B+N^i_S
-N^i_{d}+ 
N^i_{d}\log{\frac{N^i_{d}}{N^i_B+N^i_S}} \right ]
\notag 
\\
& &
+\left(\frac{\xi_b^{st}}{\sigma^{st}_b}\right)^2
+\left(\frac{\xi_b^{sy}}{\sigma^{sy}_b}\right)^2
+\left(\frac{\xi_s^{st}}{\sigma^{st}_s}\right)^2+\left(\frac{\xi_s^{sy}}
{\sigma^{sy }_s}\right)^2 \Big\}\nonumber \\
&&\equiv
\chi^2_{\rm ATLAS} (M_{Z'},G,\Gamma_{Z'})
\label{eq:chi2_atlas}
\end{eqnarray}
where
\begin{eqnarray}
N^i_B&=&N^i_b\left(1+\xi_b^{st}+\xi_b^{sy}\right) 
+N^i_{ww}\left(1+\xi_s^{st}+\xi_s^{sy}\right)\\
N^i_S&=&\left(G^2\,N^i_{Z^{\prime}}+ G \,N^i_{int}\right)
\left(1+\xi_s^{st}+\xi_s^{sy}
\right)
\label{eq:bin_atlas}
\end{eqnarray}
and $N^i_b$ is the number of background events expected in the $i$-th
bin for the SM processes except for the $W^+W^-$ contribution,
$N^i_{ww}$ stands for the number of events expected on the $i$-th bin
for the SM $W^+W^-$ contribution, and $G^2N^i_{Z^{\prime}}$ and
$GN^i_{int}$ are the number of events expected on the $i$-th bin for
the pure signal contribution and the interference
respectively. 
\smallskip

In constructing the log-likelihood function in
Eq.~\eqref{eq:chi2_atlas} we estimated the effect of the systematic
uncertainties by means of a simplified treatment in terms of four
pulls $\xi$ \cite{pulls}, where $\xi_b^{st}$ is the pull to account
for the statistical uncertainty on the evaluations for all the SM
processes except for the $W^+W^-$ contribution, $\xi_b^{sy}$ is the
one to account for the systematic uncertainty in the same processes,
$\xi_s^{st}$ is the pull to account for the statistical uncertainty on
the expectations for $W^+W^-$ and the $Z^\prime$ new contributions and
finally $\xi_s^{sy}$ accounts for the systematic uncertainty on the
same processes. The standard deviations for these pulls are obtained
from Table 6 of~\cite{ATLASww}:
\begin{eqnarray}
\sigma^{st}_b&=&0.038\ \ \ \ \ \ \sigma^{sy}_b=0.16\\
\sigma^{st}_s&=&0.0039\ \ \ \ \ \ \sigma^{sy}_s=0.093
\label{eq:pulls_atlas}
\end{eqnarray}

We performed two analyses. In the first one we computed the $\ln
L_{\rm ATLAS}$ with the 15 transverse mass bins in~\cite{ATLASww}
between $M_T=40$ GeV and $M_T=340$ GeV (\ie\ $N^{max}_{AT}=15$).  In
the second one we added an extra 16th bin (\ie\ $N^{max}_{AT}=16$)
where we sum the $Z^\prime$ expected contributions with $M_T>340$ GeV
and we assumed that the number of observed events and SM expected
predictions for the 16th bin are null.
\smallskip

\medskip
\noindent{\bf CMS analysis}
\medskip

Similarly we tuned our Monte Carlos to simulate the CMS results, by
comparing them with the CMS simulation for the SM $W^+W^-$ production
in the $ee$, $e\mu$, and $\mu\mu$ channels presented in
Ref.~\cite{CMSww}. For that we applied the selection described in
Section 3 of this reference.  In particular electrons and muons are
accepted if
\begin{eqnarray}
 &&|\eta_e|<2.5 \hbox{ and }|\eta_\mu|<2.4.
\label{cuts:cms1}
\end{eqnarray}
Also, the lepton isolation requirement in ME+Pythia+PGS-MC simulation
is that the sum of $p_T$ of all other tracks is less than 10\% of the
$p_T$ of the lepton within a cone $\Delta R <0.4~ (0.3) $ around the
electron (muon).  To implement this requirement in OUR ME-MC we simply
impose:
\begin{eqnarray}
&&\Delta R_{ee}>0.4 \hbox{ and }\Delta R_{e\mu,\mu\mu}>0.3\ 
\label{cuts:cms2}
\end{eqnarray}
Events are selected if they verify that:
\begin{eqnarray}
&&p_T^{\rm leading} > 20 \hbox{ GeV,}
\nonumber
\\
&&p_T^{\rm subleading} > 10 \hbox{ GeV,}
\nonumber
\\
&&M_{\ell\ell} > 12 \hbox{ GeV }\;\;\hbox{and}\;\; M_{e\mu} > 12 \hbox{ GeV,}
\label{cuts:cms3}
\\
&&| M_{\ell\ell} -M_Z| > 15 \hbox{ GeV, }
\nonumber
\\
&&E_{T,~rel}^{miss}(ee,\mu\mu)>40\hbox{ GeV } \;\;\hbox{and}\;\;E_{T,~rel}^{miss}(e\mu)>20\hbox{ GeV.}
\nonumber
\end{eqnarray}
In ME+Pythia+PGS-MC simulation jets are reconstructed with the anti-$k_T$
algorithm with a jet resolution parameter $\Delta R=0.5$ and we veto
events containing jets with
\begin{eqnarray}
&&  p_T>30 \hbox{ GeV  and  } |\eta_j|<5.0 \; .
\label{cuts:cms4}
\end{eqnarray}
Finally for events with same flavour leptons, the angle in the
transverse plane between the dilepton system and the most energetic
jet with $p_T>15$ GeV is required to be smaller than 165
degrees. 
\smallskip

\begin{figure}
\includegraphics[width=0.5\textwidth]{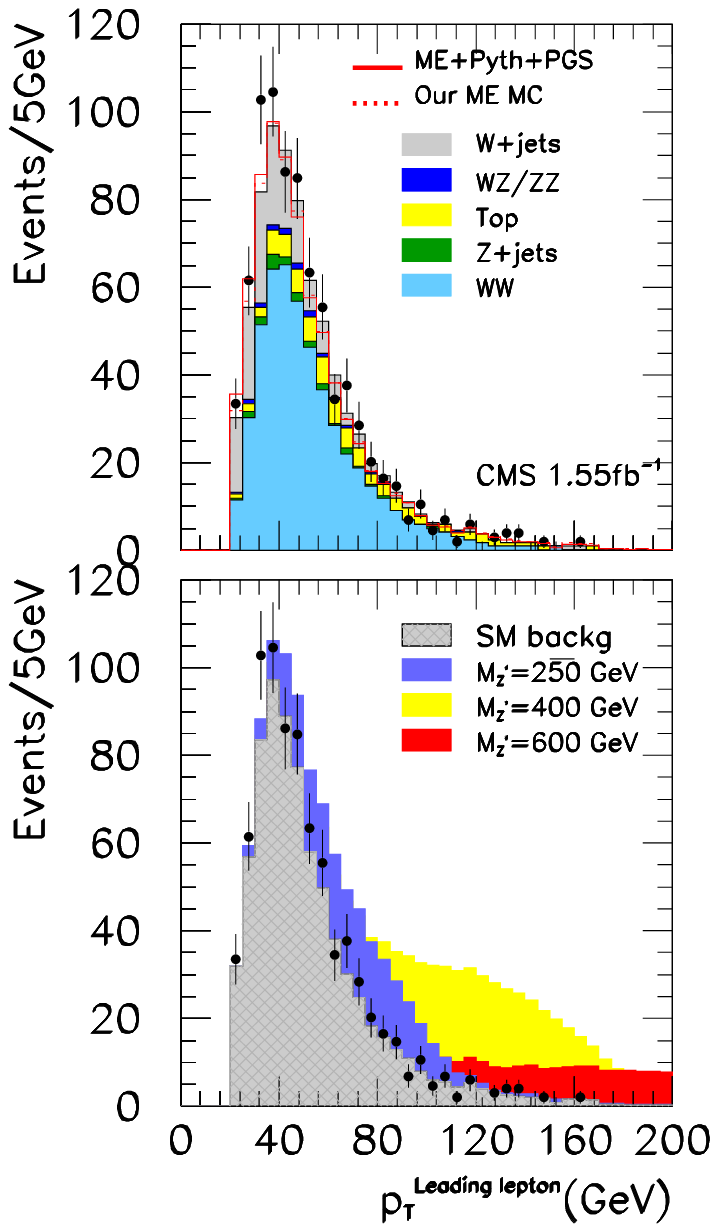}
\caption{Upper panel: Leading lepton transverse momentum distribution of the SM
  contributions to the process $pp \to \ell^+ \ell^{\prime
    -}\sla{E}_T$ calculated by CMS (colored histograms) together with
  the number of observed events by CMS (points with error bars) 
  and the performance of  ME+Pythia+PGS-MC (red solid) and 
  OUR ME-MC (red dotted). 
The results shown correspond to an integrated
luminosity of $\mathcal{L}=1.55\ \mbox{fb}^{-1}$.
\\
  Lower panel: Transverse momentum of the leading lepton for the
  total SM contribution to the process $pp \to \ell^+ \ell^{\prime
    -}\sla{E}_T$ (gray hashed) together with the total expected number
  of events   including a $Z^\prime$ of 250 GeV with $G=0.5$  (blue), a
  $Z^\prime$ of 400 GeV with 
  $G=1$ (yellow) and a $Z^\prime$ of 600 GeV with $G=1$ (red). 
For the three masses  $\Gamma_{Z^\prime}=0.06M_{Z^\prime}$. 
We include also the observed distribution of events in CMS.  }
\label{fig:dist_cms}
\end{figure}

We exhibit in Table~\ref{norma} the overall normalization needed to
tune our simulations to the CMS one presented in Table 1 of
Ref~\cite{CMSwwold}.  To verify the quality of our simulations we
compare their results with the kinematic distributions in
Ref.~\cite{CMSww}.  As an illustration in the top panel of
Figure~\ref{fig:dist_cms} we plot the leading lepton transverse
momentum distribution.  As we can see, our simulation tools are in
good agreement with the CMS Monte Carlo.  
\smallskip

As before, in the simulation of the $Z^\prime$ signal we employed the
same normalization factors obtained from the $W^+W^-$ SM production
for the channels $ee$, $e\mu$, and $\mu\mu$. Here, the presence of a
new spin--1 resonance leads to an enhancement at large $p_T$'s as
displayed in the lower panel of Fig.~\ref{fig:dist_cms}. 
\smallskip

The exclusion limits on the production of a $Z^\prime$ were extracted
using a binned log-likelihood function based on the contents of the
bins of the transverse momentum distribution of the leading
lepton\footnote{ With in the range of the kinematic variables
  presented in the different CMS plots, the leading lepton transverse
  momentum distribution is the most sensitive to the presence of a
  $Z'$.}
\begin{eqnarray}
&&-2\ln L_{\rm CMS}(M_{Z'},G,\Gamma_{Z'})= 
 \nonumber  \\
&& \begin{array}{c}\\ [-0.2cm]
{\rm Min}\\[-0.2cm] \xi_j 
\end{array}\Big\{
2\sum_{i=1}^{N^{max}_{CMS}} \left [ N^i_B+N^i_S
-N^i_{d}+
N^i_{d}\log{\frac{N^i_{d}}{N^i_B+N^i_S}} \right ]
+
\notag
\\
& &
+\left(\frac{\xi_b^{sy}}{\sigma^{sy}_b}\right)^2
+\left(\frac{\xi_s^{sy}}{\sigma^{sy}_s}\right)^2 \Big\}
\equiv 
\chi^2_{\rm CMS}(M_{Z'},G,\Gamma_{Z'})
\label{eq:chi2_cms}
\end{eqnarray}
where
\begin{eqnarray}
N^i_B&=&N^i_b\left(1+\xi_b^{sy}\right)
+N^i_{ww}\left(1+\xi_s^{sy}\right)\\
N^i_S&=&\left(G^2 N^i_{Z^{\prime}}+
G N^i_{int}\right)\left(1+\xi_s^{sy}\right) 
\; .
\label{eq:bin_cms}
\end{eqnarray}
Again $N^i_b$ stands for the number of events expected on the $i$-th bin
for the SM processes except for the $W^+W^-$ contribution, $N^i_{ww}$
is the number of events expected on the $i$-th bin for the $W^+W^-$
contribution, $G^2N^i_{Z^{\prime}}$ and $G N^i_{int}$ are the number of
events expected on the $i$-th bin for the pure signal contribution and
the interference respectively and $N^i_d$ is the observed events on
the bin $i$.  
\smallskip

In the CMS case we make a simplified treatment of the systematic
uncertainties in terms of two pulls: $\xi_b^{sy}$ is the pull to
account for the uncertainty on the expectations for all the SM
processes except for the $W^+W^-$ contribution while $\xi_s^{sy}$ is
the one to account for the systematic uncertainty on $W^+W^-$ and the
$Z^\prime$ new contributions. The standard deviations for these pulls
are obtained from~\cite{CMSwwold}:
\begin{eqnarray}
\sigma^{sy}_b&=&0.20 \; , \\
\sigma^{sy}_s&=&0.08 \; .
\label{eq:pulls_cms}
\end{eqnarray}
As for ATLAS we performed two analyses. In the first one we calculate
$\ln L_{\rm CMS}$ with the event rates in the 36 leading transverse
momentum bins between 20 GeV and 200 GeV (\ie\ $N^{max}_{CMS}=36$). In
the second analysis we included an extra bin where we sum expected
contributions from the $Z^\prime$ with $p_T^{\rm leading}>200$ GeV
(\ie\ $N^{max}_{CMS}=37$) and where we assumed that the number of
observed events and SM expected predictions for the 37th bin are equal
to 0.
\smallskip

\medskip
\noindent{\bf Combined Analysis}
\medskip

We also combined the ATLAS and CMS results to get more stringent
exclusion limits on the production of a $Z^\prime$ by constructing the
combined log-likelihood function
\begin{eqnarray}
\chi^2_{\rm comb}
(M_{Z'},G,\Gamma_{Z'})
&=&\chi^2_{\rm ATLAS}(M_{Z'},G,\Gamma_{Z'})  \nonumber \\
&&+\chi^2_{\rm CMS}(M_{Z'},G,\Gamma_{Z'})
\label{eq:chi2_combined}
\end{eqnarray}
where we conservatively assumed that the ATLAS and CMS systematic
uncertainties are uncorrelated.  
\smallskip

In all cases  we set the exclusion 95\% (2$\sigma$, 1 d.o.f) limits 
on $G$ by maximizing the corresponding likelihood function (or 
equivalently minimizing the $\chi^2$) with respect to $G$ 
for each value of $M_{Z'}$ and $\Gamma_{Z'}$ 
and imposing
\begin{eqnarray}
|\chi^2 (M_{Z'},G,\Gamma_{Z'})-\chi^2_{\rm min}(M_{Z'},\Gamma_{Z'})|>4 \; .
\end{eqnarray}

\section{Model Independent Results}
\label{sec:results}

\begin{figure*}
\includegraphics[width=0.7\textwidth]{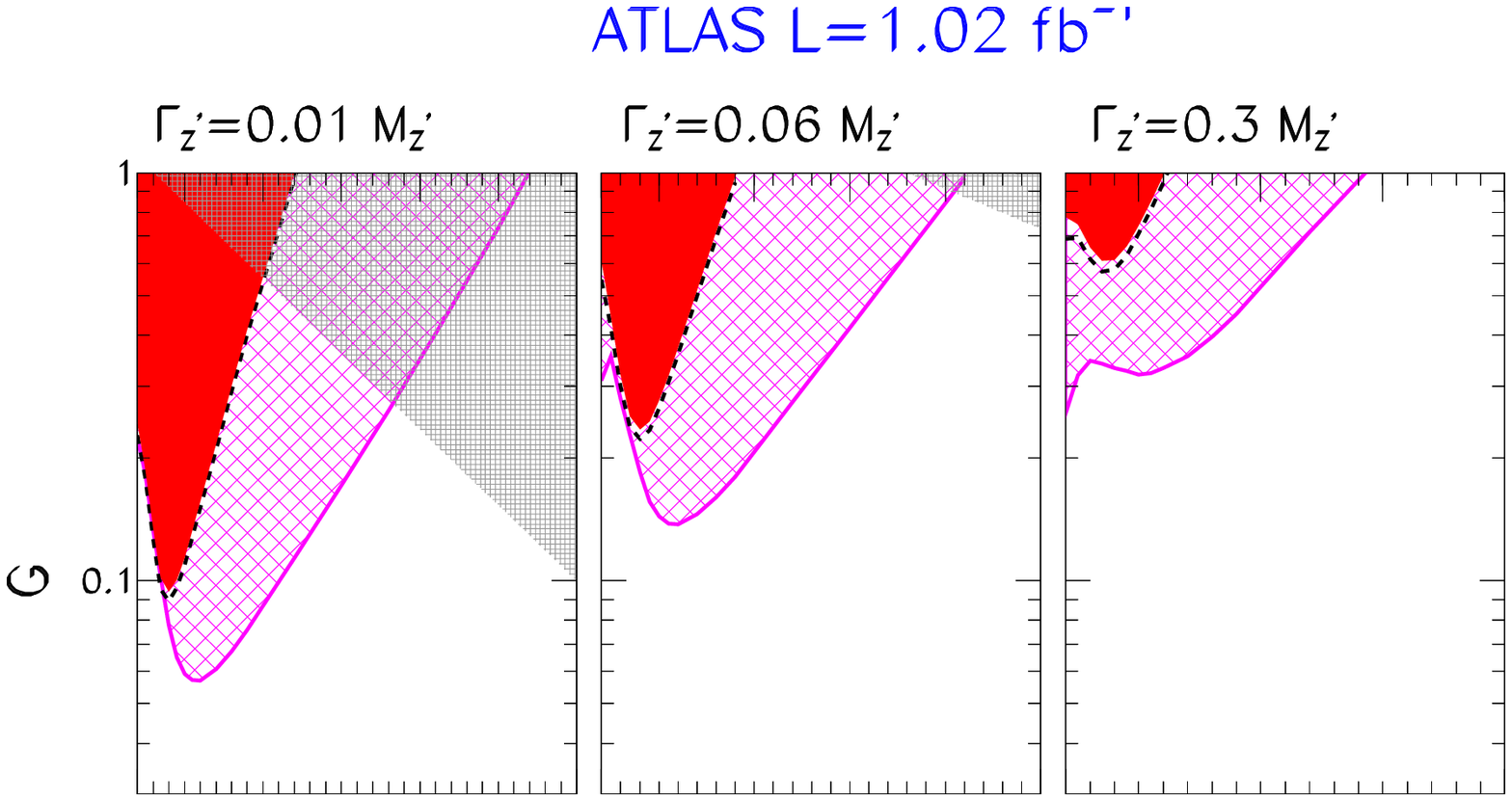}
\vglue -0.65cm
\includegraphics[width=0.7\textwidth,angle=180]{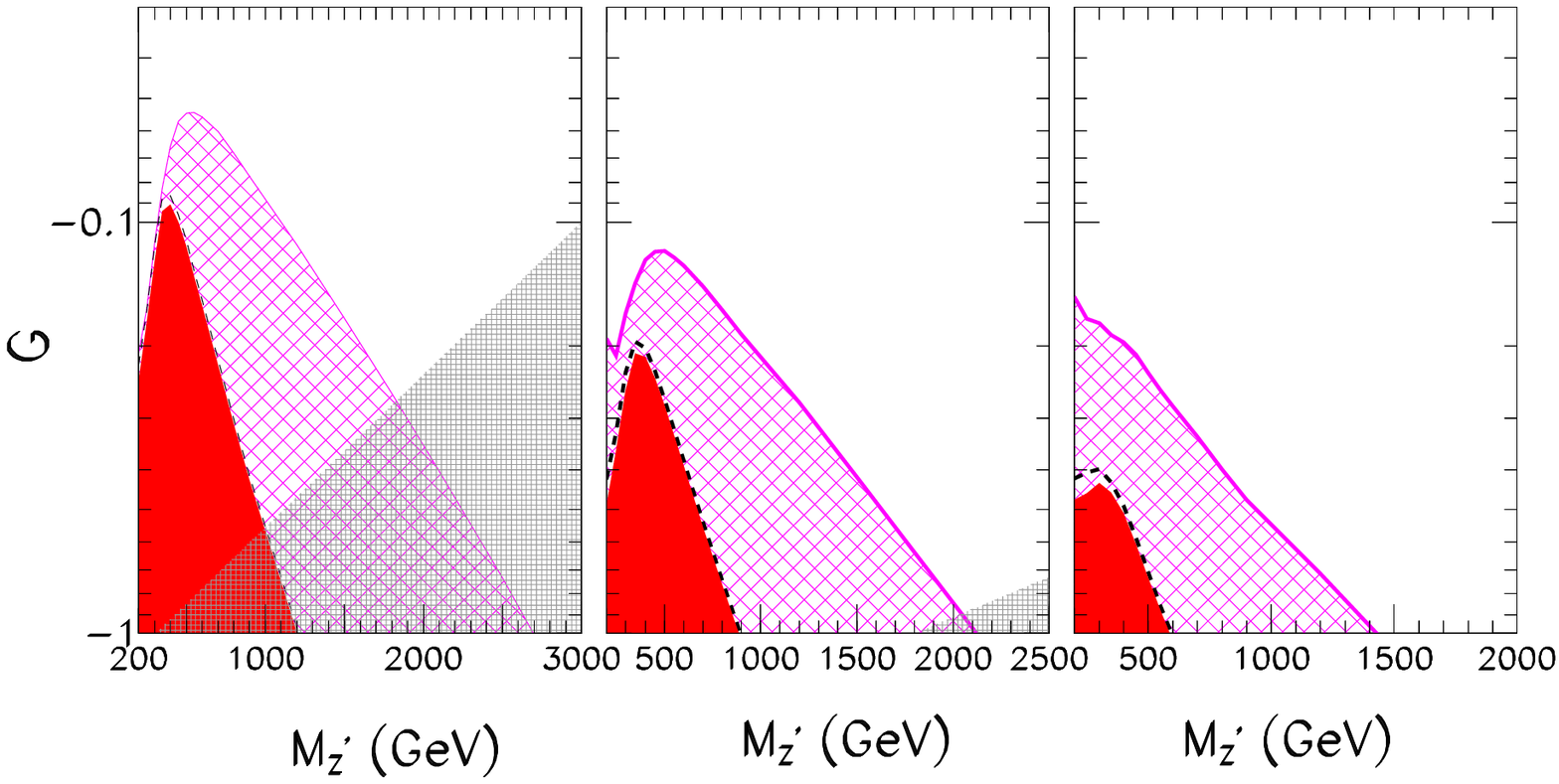}
\caption{ 95\% CL exclusion limits on the production of a $Z^\prime$
  from our analysis of the $M_T$ distribution measured by ATLAS with
  $\mathcal{L}=1.02\ \mbox{fb}^{-1}$ and for three values of
  $\Gamma_{Z'}/M_{Z'}=0.01$, $0.06$ and $0.3$ (left, center and right
  panels respectively).  The red solid regions are derived using the
  log-likelihood function in Eq.~\eqref{eq:chi2_atlas} with
  $N^{max}_{AT}=15$.  The regions bounded by the black dashed curves
  correspond to the same analysis removing the effect of the
  systematic pulls.  The purple hatched regions are derived using the
  log-likelihood function in Eq.~\eqref{eq:chi2_atlas} with
  $N^{max}_{AT}=16$.  The shadowed regions in the upper (lower) right
  corner of the upper (lower) panels represent the excluded values by
  the condition Eq.~\eqref{eq:zcouplimit}.}
\label{fig:bounds_atlas}
\end{figure*}

The $2\sigma$ exclusion limits on possible new states $Z^\prime$
derived from $M_T$ spectrum observed at the $\mathcal{L}=1.02\
\mbox{fb}^{-1}$ ATLAS data set are depicted in
Fig.~\ref{fig:bounds_atlas}. The results are shown in the plane $G
\otimes M_{Z^\prime}$ for three possible values of the $Z^\prime$
width $\Gamma_{Z'}/M_{Z'}=0.01$, $0.06$ and $0.3$ as labeled in this
figure. 
\smallskip

The red solid regions in Fig.~\ref{fig:bounds_atlas} were derived
using the log-likelihood function in Eq.~\eqref{eq:chi2_atlas} with
$N^{max}_{AT}=15$, {\em i.e.}  with the 15 bins of the transverse mass
distribution between $M_T=40$ GeV and $M_T=340$ GeV.  Comparing the
left, central and right panels one observes that, as expected, bounds
are stronger for narrow resonances.  The shadowed regions in the upper
(lower) right corner of the upper (lower) panels of this figure
represents the excluded values by the condition
Eq.~\eqref{eq:zcouplimit}. 
\smallskip

In order to illustrate the effect of the systematic uncertainties
included in this analysis we also show the black dashed curves which
correspond to the same analysis but fixing the pulls to zero.  As seen
by comparing the dashed curve with the boundary of the solid region,
the bounds are dominated by statistics for the available integrated
luminosity and the inclusion of the systematic uncertainties have a
very limited impact. 
\smallskip

The sensitivity reach when a non-zero observation for $M_T>340$ GeV is
included as a 16th bin, is shown as the purple hatched regions. The
effect of the inclusion of this additional bin is more important the
heavier and the wider $Z'$ is. This is due to the fact that a heavier
and/or wider $Z^\prime$ gives a larger contribution to events with
$M_T>340$ GeV. Finally the difference between the regions in the upper
and lower panels arises from the interference between the SM and $Z'$
contribution. As expected this effect is only relevant for the lighter
and wider $Z'$ since the interference term is roughly proportional to
$\Gamma_{Z^\prime}/M_{Z^\prime}$.  
\smallskip

\begin{figure*}
\includegraphics[width=0.7\textwidth]{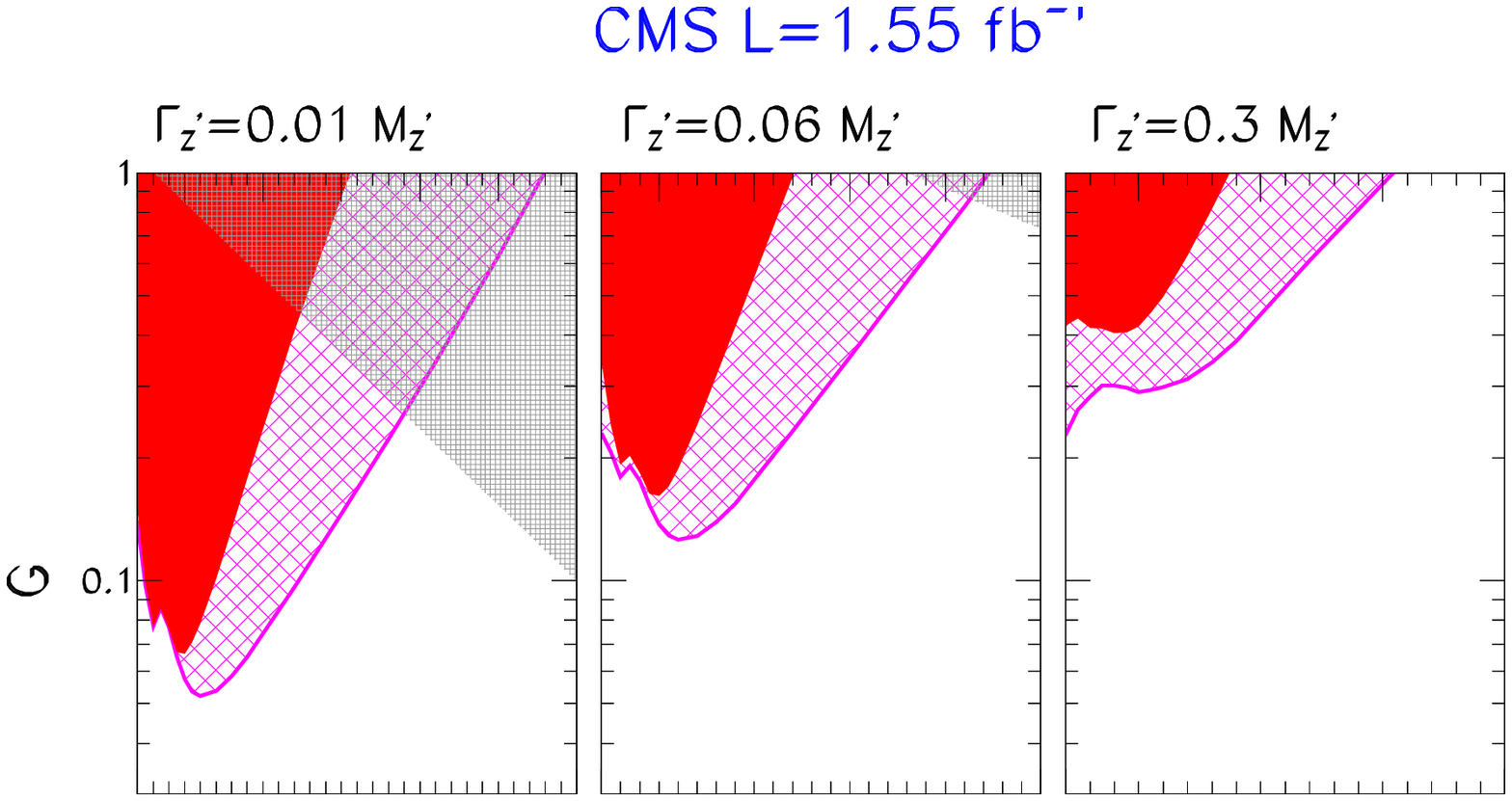}
\vglue -0.6cm
\includegraphics[width=0.7\textwidth,angle=180]{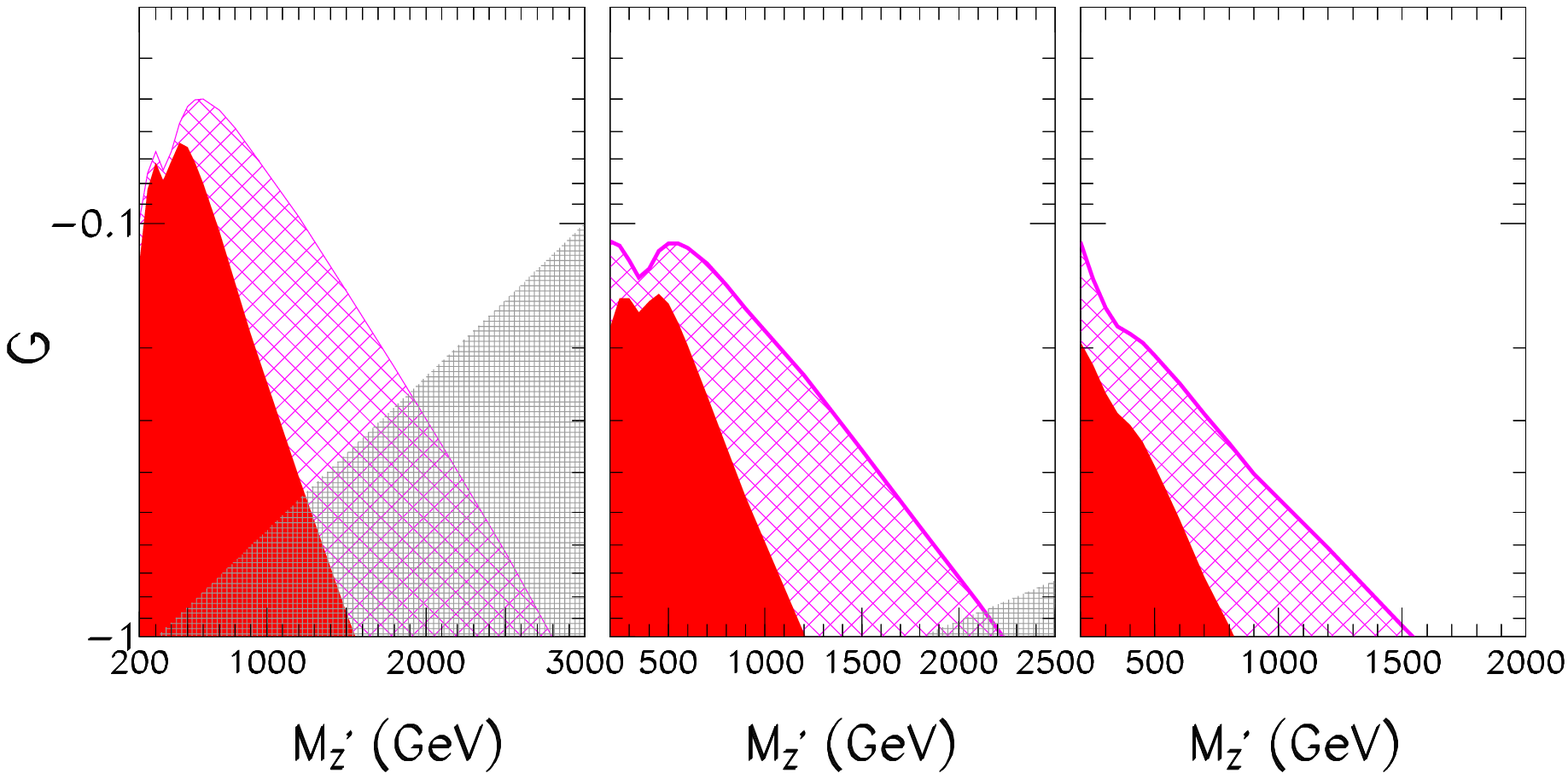}
\caption{ 95\% CL exclusion limits on the production of a $Z^\prime$
  from our analysis of the $p_T^{\rm leading}$ distribution measured
  by CMS with $\mathcal{L}=1.55\ \mbox{fb}^{-1}$.  The left, center
  and right panels correspond to three values of
  $\Gamma_{Z'}/M_{Z'}=0.01$ ,$0.06$ and $0.3$ respectively.  The red
  solid regions are derived using the log-likelihood function in
  Eq.~\eqref{eq:chi2_cms} with $N^{max}_{CMS}=36$.  The purple hatched
  regions are derived using the log-likelihood function in
  Eq.~\eqref{eq:chi2_cms} with $N^{max}_{CMS}=37$.  The shadowed
  regions in the upper (lower) right corner of the upper (lower)
  panels represent the excluded values by the condition
  Eq.~\eqref{eq:zcouplimit}.}
\label{fig:bounds_cms}
\end{figure*}

The $2\sigma$ exclusion limits on the production of a $Z^\prime$
derived from our analysis of the $p_T^{\rm leading}$ distribution
measured by CMS with $\mathcal{L}=1.55\ \mbox{fb}^{-1}$ can be seen in
Fig.~\ref{fig:bounds_cms}.  The dependence of the excluded range of
$G$ on the $Z'$ mass and width is similar to
Fig.~\ref{fig:bounds_atlas} as expected.  The only difference is
associated with the larger event sample.  As no positive signal is
observed neither in ATLAS nor in CMS, the bounds obtained from our
analysis of the CMS data are stronger than for the ATLAS due to the
larger integrated luminosity used in the former. 
\smallskip

\begin{figure*}
\includegraphics[width=0.7\textwidth]{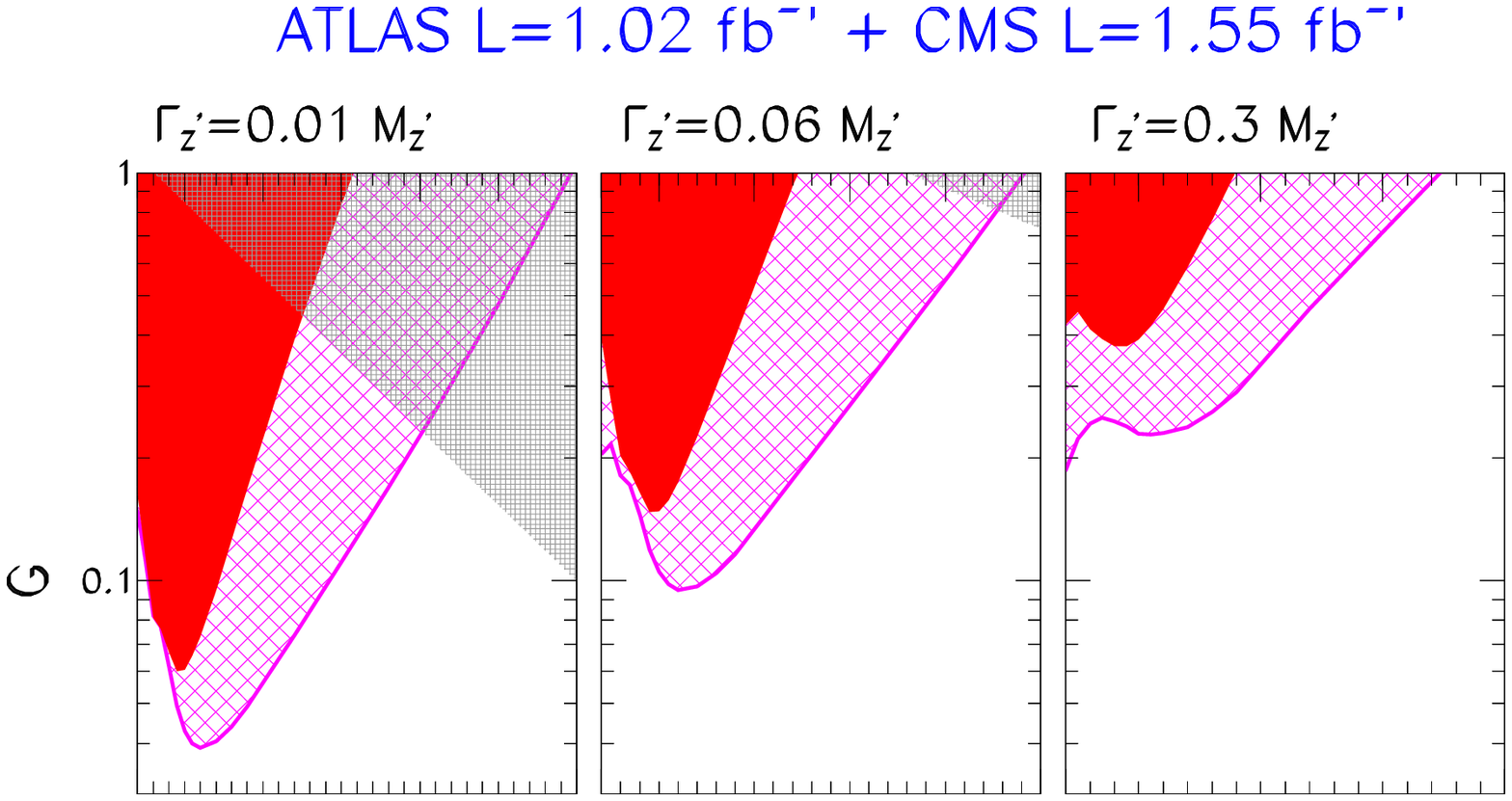}
\vglue -0.6cm
\includegraphics[width=0.7\textwidth,angle=180]{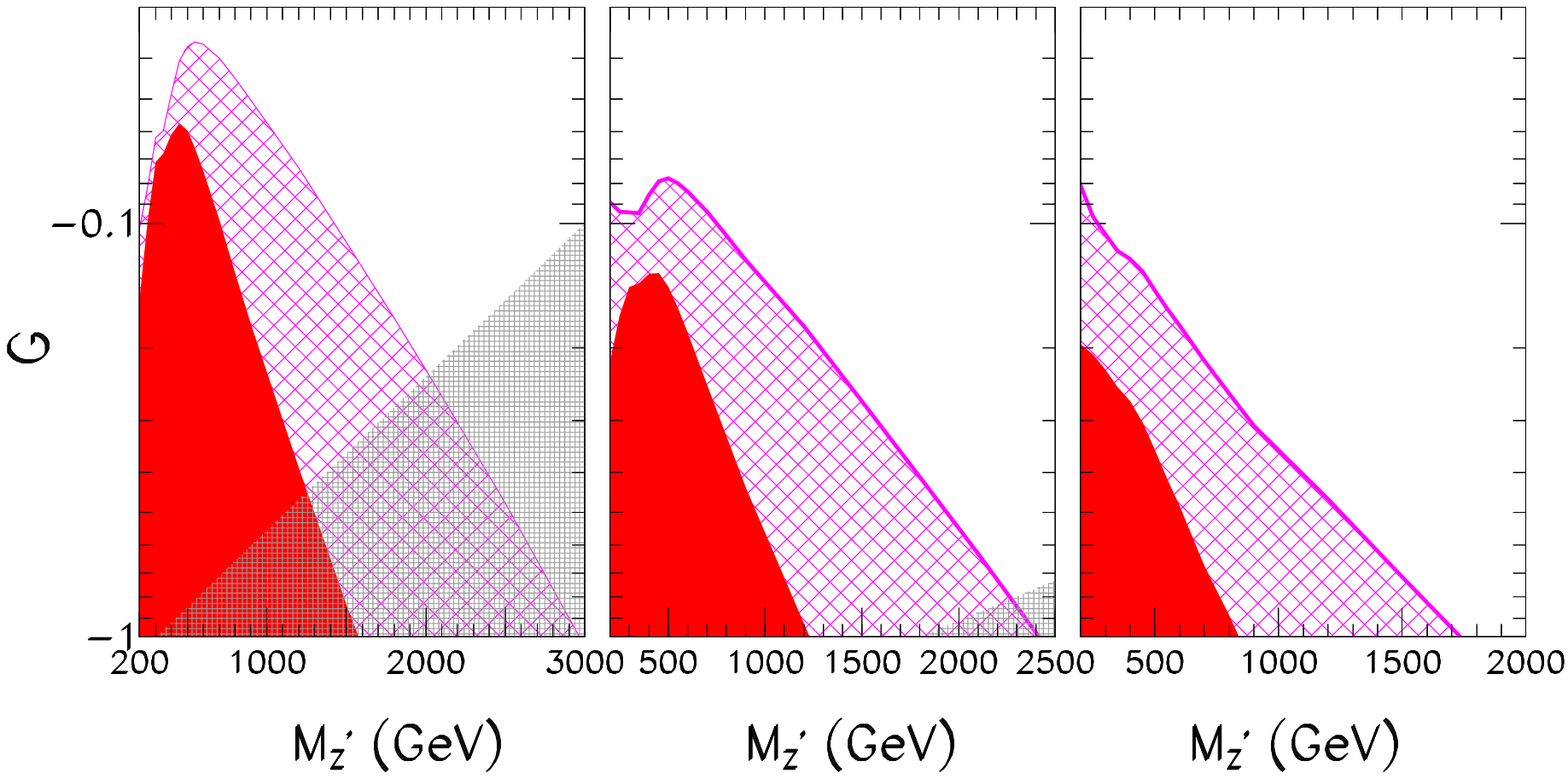}
\caption{95\% CL exclusion limits on the production of a $Z^\prime$
  from our combined analysis of the measured $M_T$ distribution in
  ATLAS with $\mathcal{L}=1.02\ \mbox{fb}^{-1}$ and the $p_T^{\rm
    leading}$ distribution measured by CMS with $\mathcal{L}=1.55\
  \mbox{fb}^{-1}$.  The red solid (purple hatched) regions are derived
  using the log-likelihood defined in Eq.~\eqref{eq:chi2_combined}
  with 15 and 36 (16 and 37) bins of the ATLAS and CMS distributions
  respectively.  The shadowed regions in the upper (lower) right
  corner of the upper (lower) panels represent the excluded values by
  the condition Eq.~\eqref{eq:zcouplimit}.}
\label{fig:bounds_comb}
\end{figure*}

Finally in Fig.~\ref{fig:bounds_comb} we present the exclusion
constraints on the production of a new neutral vector resonance from
our combined analysis of the measured $M_T$ distribution in ATLAS with
$\mathcal{L}=1.02\ \mbox{fb}^{-1}$ and the $p_T^{\rm leading}$
distribution measured by CMS with $\mathcal{L}=1.55\ \mbox{fb}^{-1}$.
We see that the combination of ATLAS and CMS data have already
excluded a sizable region of the parameter space for the production of
new spin-1 $Z^\prime$ associated with the EWSB sector.  In particular,
from our analysis with 15 and 36 (16 and 37) bins of the ATLAS and CMS
distributions, a narrow resonance of any mass with
$\Gamma_{Z'}/M_{Z'}=0.01$ and that saturates the partial wave
amplitude for the process $W^+ W^- \to W^+ W^-$, is excluded at 95\%
CL if its coupling to the light quarks is larger than 45\% (22\%) of
the SM $Z\bar{q}q$ coupling. Moreover, our analysis with 15 and 36
bins of the ATLAS and CMS distributions, excludes at 95\% CL a wider
resonance with $\Gamma_{Z'}/M_{Z'}=0.06$ $(0.3)$ that saturates the
partial wave amplitude for the process $W^+ W^- \to W^+ W^-$ and
couples to light quarks with SM strength if $M_{Z'}\leq 1250 \, (850)$
GeV. From the extended analysis using 16 and 37 bins of the ATLAS and
CMS distributions we find that no such SM coupling resonance is
allowed for any mass for $\Gamma_{Z'}/M_{Z'}=0.06$ or $M_{Z'}<1750$
GeV for $\Gamma_{Z'}/M_{Z'}=0.3$. 
\smallskip

At this point it is interesting to compare our $Z^\prime$ bounds with
the ones obtained by the CDF collaboration analyzing $WW$ production
at the Tevatron~\cite{cdf} in the framework of the Sequential Standard Model 
\cite{SSM}. In the CDF analysis our coupling $G$
is related to the parameter $\xi$ as 
$G=\xi\sqrt{3} M_{Z'}/M_Z$ while the $Z^\prime$ width is
a well defined function of $\xi$ and $M_{Z'}$. Generically 
this lead to a narrow $Z^\prime$s with $\Gamma_{Z^\prime}/M_{Z^\prime}
\lesssim 0.1 $.  For $Z^\prime$ masses of 250, 600 and 950 GeV the CDF
constraints read $|G| < 0.47$, $0.27$ and $1.36$ respectively. On the
other hand our analyses without  (with) extra bins lead to bounds $|G| <
0.20$, $0.12$ and $0.60$ ($0.18$, $0.067$,$0.15$) for the same masses.  
In conclusion, translating our bounds into the model used by CDF we get that
generically the constraints from our most conservative analysis of the
ATLAS and CMS distributions, {\em i.e.} without the extra bins, extend
the CDF exclusion to couplings about a factor 2 smaller for the
accessible mass range at Tevatron $M_{Z^\prime} \lesssim
950$. Furthermore, our results also widen the accessible
$M_{Z^\prime}$ mass range. 
\smallskip

\section{Model dependent results}
\label{sec:mdr}

\begin{figure}
\includegraphics[width=0.4\textwidth]{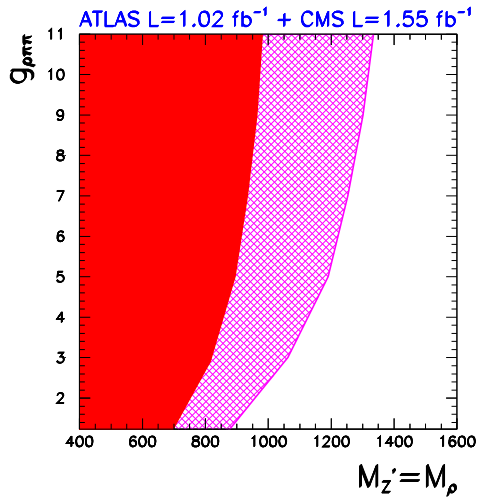}
\caption{95\% CL exclusion limits from our combined analysis of the
  measured $M_T$ distribution in ATLAS with $\mathcal{L}=1.02\
  \mbox{fb}^{-1}$ and the $p_T^{\rm leading}$ distribution measured by
  CMS with $\mathcal{L}=1.55\ \mbox{fb}^{-1}$ in the framework of the
  model in Ref.~\cite{grojean}.  The red solid (purple hatched)
  regions are derived using the log-likelihood defined in
  Eq.~\eqref{eq:chi2_combined} with 15 and 36 (16 and 37) bins of the
  ATLAS and CMS distributions respectively.}
\label{fig:grojean}
\end{figure}

The above analyses can be used to place bounds on specific models once
we take into account its couplings.  Generically within a given model
the width of the vector resonance and the strength of its couplings to
fermions and gauge bosons can be functions of a few parameters.  As an
illustration we made a dedicated study of the bounds attainable in the
framework recently proposed in Ref.~\cite{grojean} that exhibits a
single vector $SU(2)_{\rm custodial}$--triplet resonance that is
included to saturate the unitarization condition. In brief in this
case the couplings of the resonance to the fermions as well as to the
gauge bosons can be cast in terms of a unique parameter
$g_{\rho\pi\pi}$ with the decay into gauge bosons being the dominant
mode. The other free parameter is the mass of the new resonance
$M_\rho=M_{Z^\prime}$.
The limits derived in the previous section can not be directly applied to
this case since the $Z^\prime$ couplings to quarks differs from the SM
ones. In this example we generated the ${\cal O}(\alpha^4)$ amplitudes
using MADGRAPH.  The constraints in this scenario coming from the
reaction ~\ref{eq:proc} are shown in Fig.~\ref{fig:grojean} and they
represent the strongest bounds at present on this scenario. 
\smallskip

Because of the existence of an associated charged resonance
associated to the unitarization of the channel $WZ \to WZ$, bounds
can be also imposed from the searches of $pp\rightarrow Z W^\pm$ such
as the one performed by the CMS collaboration~\cite{CMSW'}. CMS
present the results of their negative searches for $W'$ in the
framework of the Sequential Standard Model~\cite{SSM} as constraints
on $\sigma(pp\rightarrow W')\times Br(W'\rightarrow 3l \nu)$.  In
Ref.~\cite{grojean} a simplified adaptation of this CMS bound was made
which seemed to exclude $M_\rho<900$ GeV for all values of
$g_{\rho\pi\pi}>1$. However one must notice that despite the bounds in
Ref.~\cite{CMSW'} are presented in a seemingly ``model independent''
form, the actual efficiency for reconstruction of their resonance
signal depends on the assumed width of the resonance which depends on
the model assumed. 
\smallskip

\section{Summary}
\label{sec:summary}

In this work we have presented an analysis of the ATLAS~\cite{ATLASww}
and CMS~\cite{CMSww} kinematic distributions of the $pp\to
\ell^+\ell^{\prime -}\sla{E}_T$ events to place bounds on the
production of a $Z^\prime$ associated with the EWSB sector which
contributes to the above final state via $ p p \to Z^\prime \to W^+
W^- \to \ell^+ \ell^{\prime -} \, \sla{E}_T$. 
\smallskip

To make our study as model independent as possible we kept as
independent parameters the coupling strength of the $Z'$ to light
quarks, to the gauge bosons, its width, and its mass. We have set
exclusion bounds by looking at the different behaviour of the SM
processes and $Z^\prime$ new contributions with respect to two
kinematical variables; the transverse momentum of the leading lepton
for the CMS case and the transverse mass of the system for the ATLAS
one as a function of the three free parameters in the study. The
results are shown in Figs.~\ref{fig:bounds_atlas} and
Figs.~\ref{fig:bounds_cms} for the study of the measured distribution
of events in ATLAS with integrated luminosity of $\mathcal{L}=1.02
\mbox{fb}^{-1}$ and in CMS with integrated luminosity of
$\mathcal{L}=1.55\ \mbox{fb}^{-1}$ respectively.  We have also
combined the likelihoods for the two analyses to get the more
stringent combined exclusion limits shown in
Fig.~\ref{fig:bounds_comb}. 
\smallskip

We observe that the combined analysis already excludes a large region
of the parameter space for the lightest masses, well exceeding the
limits from Tevatron. Moreover, we also showed how  our analysis
framework can be adapted to specific models. 
\smallskip

\section*{Acknowledgments}

We thank Sergio Novaes and Flavia Dias for clarifications about the CMS data.
O.J.P.E is supported in part by Conselho Nacional de Desenvolvimento
Cient\'{\i}fico e Tecnol\'ogico (CNPq) and by Funda\c{c}\~ao de Amparo
\`a Pesquisa do Estado de S\~ao Paulo (FAPESP); M.C.G-G is also
supported by USA-NSF grant PHY-0653342, by CUR Generalitat de
Catalunya grant 2009SGR502 and together with J.G-F by MICINN
FPA2010-20807 and consolider-ingenio 2010 program CSD-2008-0037. J.G-F
is further supported by Spanish ME FPU grant AP2009-2546.


\bibliographystyle{h-physrev4}

\begin{thebibliography}{10}

\bibitem{Lee:1977eg}
  B.~W.~Lee, C.~Quigg and H.~B.~Thacker,
  Phys.\ Rev.\  D {\bf 16}, 1519 (1977).

\bibitem{TC}
S.~Dimopoulos and L.~Susskind,
Nucl.\ Phys.\ B {\bf 155}, 237 (1979);
L.~Susskind,
Phys.\ Rev.\ D {\bf 20}, 2619 (1979);
S.~Weinberg,
Phys.\ Rev.\ D {\bf 19}, 1277 (1979).


\bibitem{NTC} See for instance,
  C.~T.~Hill and E.~H.~Simmons,
  Phys.\ Rept.\  {\bf 381}, 235 (2003)
  [Erratum-ibid.\  {\bf 390}, 553 (2004)]
  [arXiv:hep-ph/0203079];


\bibitem{hless}
C.~Csaki, C.~Grojean, H.~Murayama, L.~Pilo and J.~Terning,
Phys.\ Rev.\ D {\bf 69}, 055006 (2004)
[arXiv:hep-ph/0305237];
C.~Csaki, C.~Grojean, L.~Pilo and J.~Terning,
Phys.\ Rev.\ Lett.\  {\bf 92}, 101802 (2004)
[arXiv:hep-ph/0308038];
Y.~Nomura,
JHEP {\bf 0311}, 050 (2003)
[arXiv:hep-ph/0309189];
C.~Csaki, C.~Grojean, J.~Hubisz, Y.~Shirman and J.~Terning,
Phys.\ Rev.\ D {\bf 70}, 015012 (2004)
[arXiv:hep-ph/0310355];
G.~Cacciapaglia, C.~Csaki, G.~Marandella and J.~Terning,
  Phys.\ Rev.\  D {\bf 75}, 015003 (2007)
  [arXiv:hep-ph/0607146];
  C.~Csaki and D.~Curtin,
  arXiv:0904.2137 [hep-ph].

\bibitem{Csaki:2003dt}
  C.~Csaki, C.~Grojean, H.~Murayama, L.~Pilo and J.~Terning,
  Phys.\ Rev.\  D {\bf 69}, 055006 (2004)
  [arXiv:hep-ph/0305237].

\bibitem{ads-cft}
N.~Arkani-Hamed, M.~Porrati and L.~Randall,
JHEP {\bf 0108}, 017 (2001)
[arXiv:hep-th/0012148];
R.~Rattazzi and A.~Zaffaroni,
JHEP {\bf 0104}, 021 (2001)
[arXiv:hep-th/0012248];
M.~Perez-Victoria,
JHEP {\bf 0105}, 064 (2001)
[arXiv:hep-th/0105048].


\bibitem{Alves:2009aa}
  A.~Alves, O.~J.~P.~Eboli, D.~Goncalves {\it et al.},
  Phys.\ Rev.\  {\bf D80}, 073011 (2009).
  [arXiv:0907.2915 [hep-ph]].

\bibitem{Eboli:2011bq}
  O.~J.~P.~Eboli, C.~S.~Fong, J.~Gonzalez-Fraile, M.~C.~Gonzalez-Garcia,
  Phys.\ Rev.\  {\bf D83}, 095014 (2011).
  [arXiv:1102.3429 [hep-ph]].


\bibitem{Birkedal:2004au}
  A.~Birkedal, K.~Matchev and M.~Perelstein,
  Phys.\ Rev.\ Lett.\  {\bf 94}, 191803 (2005)
  [arXiv:hep-ph/0412278];
  H.~J.~He {\it et al.},
  Phys.\ Rev.\  D {\bf 78}, 031701 (2008)
  [arXiv:0708.2588 [hep-ph]];
  T.~Ohl and C.~Speckner,
  Phys.\ Rev.\  D {\bf 78}, 095008 (2008)
  [arXiv:0809.0023 [hep-ph]].

\bibitem{ATLASww}
  The ATLAS Collaboration,
  ATLAS-CONF-2011-110

\bibitem{CMSww}
  The CMS Collaboration
  CMS-HIG-11-014
 \begin{verbatim}
 https://twiki.cern.ch/twiki/bin/view
/CMSPublic/Hig11014TWiki
 \end{verbatim}

\bibitem{madevent}
T.~Stelzer and F.~Long,
Comput.{} Phys.{} Commun.{} \textbf{81} (1994) 357;
F.~Maltoni and T.~Stelzer,
J.\ High Energy Phys. {\bf 0302}, 027 (2003)
[arXiv:hep-ph/0208156].


\bibitem{Sjostrand:2006za}
  T.~Sjostrand, S.~Mrenna, P.~Z.~Skands,
  JHEP {\bf 0605}, 026 (2006).
  [hep-ph/0603175].


\bibitem{pgs4} John Conway, PGS 4, 
 \begin{verbatim}
 https://physics.ucdavis.edu/~conway/research
/software/pgs/pgs4-support.htm
 \end{verbatim}

\bibitem{CTEQ6}
J.~Pumplin, D.~R.~Stump, J.~Huston, H.~L.~Lai, P.~Nadolsky and W.~K.~Tung,
  JHEP {\bf 0207}, 012 (2002)
  [arXiv:hep-ph/0201195].

\bibitem{arXiv:0802.1189}
  M.~Cacciari, G.~P.~Salam and G.~Soyez,
  JHEP\ {\bf 0804} (2008) 063
  [arXiv:0802.1189 [hep-ph]].


\bibitem{pulls}
  G.~L.~Fogli, E.~Lisi, A.~Marrone, D.~Montanino, A.~Palazzo,
  Phys.\ Rev.\  {\bf D66}, 053010 (2002).
  [hep-ph/0206162],
 M.~C.~Gonzalez-Garcia, M.~Maltoni,
  Phys.\ Rept.\  {\bf 460}, 1-129 (2008).
  [arXiv:0704.1800 [hep-ph]].


\bibitem{CMSwwold}
  The CMS Collaboration
  CMS-EWK-11-010

\bibitem{cdf}
T.~Aaltonen {\it et al.} [ The CDF Collaboration ],
  Phys.\ Rev.\ Lett.\  {\bf 104}, 241801 (2010).
  [arXiv:1004.4946 [hep-ex]].

\bibitem{SSM}J.C. Pati and A. Salam,
  Phys.\ Rev.\  D {\bf 10}, 275 (1974);
Phys.\ Rev.\  D {\bf 11}, 703 (1974).

\bibitem{grojean} 
  A.~Falkowski, C.~Grojean, A.~Kaminska, S.~Pokorski and A.~Weiler,
  JHEP\ {\bf 1111}, 028  (2011)
  [arXiv:1108.1183 [hep-ph]].

\bibitem{CMSW'} 
  The CMS Collaboration
  CMS-PAS-EXO-11-041

%
%
%
%
%
%
%
%

\end{thebibliography}

\end{document}